\documentclass{article}

\usepackage{amssymb,amsmath,url,bm,enumerate}
\usepackage{booktabs, multirow}
\usepackage{graphicx,color}
\usepackage{subfigure}

\usepackage{geometry}
\title{Geometric reconstruction methods for electron tomography\footnote{NOTICE: this is the author's version of a work that was accepted for publication in \textbf{Ultramicroscopy}. Changes resulting from the publishing process, such as peer review, editing, corrections, structural formatting, and other quality control mechanisms may not be reflected in this document. Changes may have been made to this work since it was submitted for publication. A definitive version was subsequently published in Ultramicroscopy (2013), DOI: 10.1016/j.ultramic.2013.01.002}}
\author{Andreas Alpers\\
{\small Zentrum Mathematik, Technische Universit\"at M\"unchen} \\ {\small D-85747 Garching bei M\"unchen, Germany}\\
{\small \texttt{alpers@ma.tum.de}}\\
\and
Richard J. Gardner\\
{\small Department of Mathematics, Western Washington University}\\ {\small Bellingham, WA 98225-9063, USA}\\
{\small \texttt{Richard.Gardner@wwu.edu}}\\
\and
Stefan K\"onig\\
{\small Zentrum Mathematik, Technische Universit\"at M\"unchen}\\ {\small D-85747 Garching bei M\"unchen, Germany}\\
{\small \texttt{koenig@ma.tum.de}}\\
\and
Robert S. Pennington\\
{\small Center for Electron Nanoscopy, Technical University of Denmark}\\
{\small DK-2800 Kongens Lyngby, Denmark. }\\ 
{\small Now at Institut f\"ur Experimentelle Physik, Universit\"at Ulm, D-89081 Ulm, Germany}\\
{\small \texttt{robert.pennington@uni-ulm.de}}\\
\and
Chris B. Boothroyd\\
{\small Ernst Ruska-Centre for Microscopy and Spectroscopy with Electrons and Peter Gr\"unberg Institute}\\
{\small Forschungszentrum J\"ulich, D-52425 J\"ulich, Germany}\\
{\small \texttt{ChrisBoothroyd@cantab.net}}\\
\and
Lothar Houben\\
{\small Ernst Ruska-Centre for Microscopy and Spectroscopy with Electrons and Peter Gr\"unberg Institute}\\
{\small Forschungszentrum J\"ulich, D-52425 J\"ulich, Germany}\\
{\small \texttt{l.houben@fz-juelich.de}}\\
\and
Rafal E. Dunin-Borkowski\\
{\small Ernst Ruska-Centre for Microscopy and Spectroscopy with Electrons and Peter Gr\"unberg Institute}\\
{\small Forschungszentrum J\"ulich, D-52425 J\"ulich, Germany}\\
{\small \texttt{rdb@fz-juelich.de}}\\
\and
Kees Joost Batenburg\\
{\small Centrum Wiskunde \& Informatica, NL-1098XG, Amsterdam, The Netherlands}\\
{\small and Vision Lab, Department of Physics}\\
{\small University of Antwerp, B-2610 Wilrijk, Belgium}\\
{\small \texttt{Joost.Batenburg@cwi.nl}}\\
}

\begin{document}
\maketitle

\begin{abstract}
Electron tomography is becoming an increasingly important tool in materials science for studying the three-dimensional morphologies and chemical compositions of nanostructures. The image quality obtained by many current algorithms is seriously affected by the problems of \emph{missing wedge artefacts} and \emph{non-linear
projection intensities due to diffraction effects}. The former refers to the fact that data cannot be acquired over the full $180^\circ$ tilt range; the latter implies that for some
orientations, crystalline structures can show strong contrast changes. To overcome these problems we introduce and discuss several  algorithms from the mathematical fields of \emph{geometric} and \emph{discrete tomography}.  The algorithms incorporate geometric prior knowledge (mainly convexity and homogeneity), which also in principle considerably reduces the number of tilt angles required. Results are discussed for the reconstruction of an InAs nanowire.
\end{abstract}

\section{Introduction}
Missing wedge artefacts and non-linear projection intensities due to diffraction effects are known to cause severe difficulties in electron tomography (ET) reconstructions obtained by standard methods. This has been reported, e.g., in \cite{artifacts1, artifacts2, artifacts3, electrontomo09, artifacts4, artifacts5, ourpaper, artifacts6}. Nevertheless, standard methods, such as \emph{filtered backprojection}, \emph{algebraic reconstruction techniques}, and \emph{simultaneous iterative reconstruction techniques} \cite{herman, kakslaney}, are still widely used due to an apparent lack of alternatives~\cite{CTalgorithms09, micronpaper}.

However, alternatives exist in the mathematical literature. \emph{Geometric tomography} \cite{gardner}, for instance, is concerned in part with the tomographic reconstruction of homogeneous (i.e., geometric) objects. Similarly, \emph{discrete tomography} \cite{kubaherman1, kubaherman2} usually deals with objects for which atomicity is a constraint or objects that exhibit a small number of attenuation coefficients. In many applications, certain prior knowledge about the shape of the structure of interest is available. For example, when reconstructing nanorods, nanowires or certain types of nanoparticles, one can typically assume that the structures are convex. (A subset $K$ of points in the plane is \emph{convex} if for any two points in $K$ the line segment joining these two points lies completely within $K$; see Fig.~\ref{fig:support}.) In particular, in our experimental application of reconstructing an InAs nanowire from high-angle annular dark-field scanning transmission electron microscopy (HAADF STEM) data, it is known that the object is comprised of cross-sections that are mostly close to regular hexagons; see \cite{nanogrowth, hexagoncrosssection}.

Here we demonstrate the use of geometric prior knowledge to overcome the problems of missing wedge and non-linear projection intensities due to diffraction effects by introducing four algorithms. For now, we use their abbreviated names; they are introduced in the next section. One of the algorithms ($2n$-GON) appears here for the first time and uses the strongest geometric prior knowledge available in our setup, namely that the slices contain nearly regular hexagons. Two algorithms, GKXR and MPW, are introduced here for the first time in the ET context and another algorithm (DART) is applied here for the first time to the reconstruction of a nanowire. As a fifth method, we discuss the BART algorithm, which was introduced in the 1970s and performs very well on homogeneous objects. BART has been implemented in the open-source software SNARK09 \cite{snark09} and we provide commands and parameters that yield high quality reconstructions in our context.

The idea of using geometric prior knowledge in ET appears already in \cite{petersen09, electrontomo08}. In particular, \cite{electrontomo08} is, to the best of our knowledge, the only paper in ET that discusses geometric 2D slice-by-slice reconstruction methods, all of which are variants of the \emph{unfiltered backprojection} (U-FBP) algorithm. Our work takes these investigations a step further and introduces alternative reconstruction methods, some of which are mathematically proven to converge towards the solution as the number of noisy measurements tends to infinity. We compare these methods (including U-FBP and the standard method SIRT as sixth and seventh algorithms, respectively) and show that these methods perform differently depending on the experimental setup and type of noise present in the data.

The geometric reconstruction algorithms use prior knowledge to address the problem of the missing wedge and non-linear projection intensities. We show that reconstructions that are accurate (mean of the reconstructions is close to the true object) and precise (reconstructions have small variance) can be obtained by using data from considerably fewer tilt angles than in conventional tomography. This has practical consequences, because rapid data acquisition allows time resolved studies and imaging of beam-sensitive samples.

As already mentioned, we investigate here only the use of geometric prior knowledge. Other types of prior knowledge might be related to the sparsity of the signal that represents the image gradient \cite{tv1, tv2}, or it might be assumed that the object is a realization of a random process with a given probability distribution \cite{apkh-06, Gibbs1}. For further applications of special-purpose reconstruction methods in a crystallographic context, see \cite{joostnature, apkh-06, DART1, kubaherman2}.

\section{Algorithms}
In this section we briefly describe the \emph{simultaneous iterative reconstruction technique (SIRT)}, the \emph{binary algebraic reconstruction technique (BART)}, the \emph{discrete algebraic reconstruction technique (DART)}, the \emph{Gardner-Kiderlen X-ray (GKXR) algorithm}, \emph{unfiltered backprojection (U-FBP)}, the \emph{modified Prince-Willsky algorithm (MPW)}, and \emph{$2n$-GON}.  Further details on SIRT, BART, DART, GKXR, and MPW can be found in \cite{SIRT, bart2, DART2, GardnerKiderlen, gardnerkiderlen09, electrontomo08}. Our results are based on Matlab implementations of these algorithms; BART is part of the open-source  software SNARK09 \cite{snark09}.

We consider only 2D versions of each algorithm, so that 3D reconstructions are obtained by 2D slice-by-slice reconstructions. Alternative reconstruction principles exist. For 3D and 2.5D approaches employing generalized Kaiser-Bessel window functions (blobs) for not necessarily homogeneous or convex objects, see \cite{blob1,blob2}.  We assume in this paper an acquisition geometry with a single tilt axis and a limited angular range.

The algorithms require different input data. While SIRT, BART, DART, and GKXR take the projections as input, only the shadows are used in U-FBP, MPW, and $2n$-GON. Note that following (non-mathematical) standard convention, \emph{projection} refers to the measured intensity data (i.e., line integrals), while \emph{shadow} denotes their support (i.e., the detector pixel locations that record non-zero intensities). It can therefore be expected that U-FBP, MPW, and $2n$-GON are rather insensitive to intensity-affecting noise if the signal can still be distinguished from the background. Also note that the object's shadows represent binary data, which initially need to be extracted from the projection data. In this paper we achieve this by applying a suitable threshold  and filter to the projection data; see Sections~\ref{specifics1} and~\ref{specifics2}. For more advanced variants, such as edge enhancement, see \cite{electrontomo08}.

Another difference between the algorithms is that they use different types of prior knowledge. Most of the algorithms exploit the object's convexity or homogeneity. However, $2n$-GON uses the additional assumption that the object is nearly a regular $2n$-gon. A summary is given in Table~\ref{table1}.

\begin{table}[htb]
\begin{center}
{\footnotesize
\begin{tabular}{lccccc} \toprule
\textbf{Algorithm} & \multicolumn{2}{l}{\textbf{Tomographic Data}} &&\multicolumn{2}{l}{\hspace*{2ex}\textbf{Prior Knowledge}} \\
                   &  projection & shadow       && convexity & other\\ \midrule
SIRT &  \checkmark & -- &&-- & --\\ \midrule
BART & \checkmark & -- && -- & homogeneity \\ \midrule
DART & \checkmark & -- && -- & homogeneity \\ \midrule
GKXR & \checkmark & -- && \checkmark & homogeneity  \\ \midrule
U-FBP & -- & \checkmark && \checkmark& homogeneity \\ \midrule
MPW & -- & \checkmark && \checkmark & homogeneity  \\ \midrule
\multirow{2}{*}{$2n$-GON} & \multirow{2}{*}{--} & \multirow{2}{*}{\checkmark} && \multirow{2}{*}{\checkmark} & close to a \\
&&&&& regular $2n$-gon  \\ \bottomrule
\end{tabular}
}
\end{center}
\caption{Overview of algorithms.} \label{table1}
\end{table}

\subsection{Pixel-based reconstruction methods}
The first three methods discussed in this paper aim to reconstruct the individual pixel values of an image representing the object.

\subsubsection{Simultaneous iterative reconstruction technique (SIRT)} \label{sect:SIRT}
In this paper we describe and employ an additive variant of SIRT \cite[Section~7]{kakslaney}. The additive SIRT algorithm (from now on referred to as the \emph{SIRT algorithm}) is a standard technique for reconstructing grayscale images from tomographic data.
When reconstructing homogeneous objects, an additional segmentation step is required that yields a binary image.
We recall that SIRT is an iterative reconstruction algorithm that computes an approximate solution of the linear system $Ax = b$,
where the vector $x\in\mathbb{R}^n$ contains the gray level (also referred to as \emph{pixel value}) for each pixel, the vector $b\in \mathbb{R}^m$ contains the measured projection data, and the matrix $A=(a_{ij})\in\mathbb{R}^{m\times n}$
represents the projection operation (i.e., computing the product $Ax$ yields the projections corresponding to the image $x$). If no exact solution of this system
exists, SIRT computes a solution for which the norm of the difference $||Ax - b||$ (referred to as
projection error) between the computed projection and the measured data is minimal with respect to a weighted $L_2$ norm, i.e., a weighted least-squares solution (see \cite{SIRT,SIRT2} for details).
In contrast to other popular iterative algorithms, such as ART (\emph{algebraic reconstruction technique} \cite[Chapter~11]{herman}), the SIRT algorithm computes the projections for all angles in each iteration. Then the difference  between these projections and the measured projection data is computed. Subsequently, each image pixel value is updated by adding a weighted average of the projection difference for all lines that intersect this pixel. In other words,
$$
x^{(k+1)} = x^{(k)}-\lambda C A^TD(Ax^{(k)}-b),
$$
where $C=\textnormal{diag}(1/c_1,\dots,1/c_n) \in \mathbb{R}^{n{\times}n}$ with $c_j=\sum_{i=1}^ma_{ij}$ and $D=\textnormal{diag}(1/d_1,\dots,1/d_m) \in \mathbb{R}^{m{\times}m}$ with $d_i=\sum_{j=1}^na_{ij}$; the parameter $\lambda \in \mathbb{R}$ is the relaxation parameter of the algorithm (for our particular choice, see Sections~\ref{specifics1} and~\ref{specifics2}).

\subsubsection{Binary algebraic reconstruction technique (BART)}
The BART algorithm was introduced by Herman \cite{bart2}. Along with other methods, BART is implemented in the open-source software SNARK09 \cite{snark09}. The general idea is to enforce binary constraints on the solution $x$ during the iterations of a chosen ART routine. In a second step, the solution is filtered to exclude isolated pixels. A complete code that can be read into SNARK09 (either manually or as an input file) to obtain the BART reconstructions discussed in this paper is provided in Appendix~A (Table~\ref{table:BART}).

\subsubsection{Discrete algebraic reconstruction technique (DART)}
The DART algorithm, which has recently been proposed as a reconstruction algorithm for electron tomography \cite{DART1,DART2} is another algebraic method, in which a set of \emph{fixed pixels} is updated in each iteration, reflecting the position of the boundary in the current reconstruction. The variant described here does not apply any subsequent filtering. For simplicity, we describe the DART algorithm for the specific purpose of reconstructing a binary image. The general algorithm can deal with more than two gray levels. More on DART can be found in \cite{DART4,DART3,DART6,DART5}.

Here, we focus on reconstructing a single object of homogeneous composition, resulting in a binary image reconstruction problem.
The DART algorithm uses a continuous algebraic reconstruction method, such as SIRT, as a subroutine. From this point on, we will refer to this continuous method as the \emph{algebraic reconstruction method (ARM)}. After an initial gray level reconstruction has been computed using the ARM, this gray level reconstruction is segmented by global thresholding with threshold $\rho/2$, where the gray level $\rho$ of the object is assumed to be prior knowledge. In practice, an appropriate value of $\rho$ can often be obtained by first computing a SIRT reconstruction and then taking the average gray level over a region deeply in the interior of the object. One of the principal assumptions behind the DART algorithm is that errors in this segmentation are typically located near the boundary of the structure of interest. Indeed, when reconstructing a homogeneous object that is large with respect to the image pixel size, pixels that are deeply inside the interior of the object (e.g., a nanoparticle) are usually segmented correctly, while the segmentation of the boundary can be highly inaccurate, in particular when the ARM reconstruction suffers from missing wedge artefacts. The boundary can be computed from the initial segmentation as the set of pixels for which not all neighboring pixels belong to the same segmentation class. After the segmentation step, the set of pixels is separated into three subsets: the interior pixels $I$, the background pixels $B$ and the boundary pixels $F$. Prior knowledge about the gray level $\rho$ of the object and of the gray level of the background (here assumed to be $0$) is now incorporated by solving the following constrained reconstruction problem, again using the ARM:

$$
\begin{array}{l}
   \mathbf{solve}\quad Ax = b\\
   \mathbf{subject\,\,to}\\
   \begin{array}{ll}
   x_i = 0, &\mathrm{for}\quad i \in B,\\
   x_i = \rho, &\mathrm{for}\quad i \in I.\\
   \end{array}
\end{array}
$$

In this reconstruction problem, all interior pixels and background pixels are fixed to their respective gray levels and pixel values are only allowed to change for boundary pixels. This constraint strongly reduces the number of unknowns in the equation system, while the number of equations (i.e., the number of entries in $b$) remains unaltered. If the initial segmentation is of sufficient quality, the reconstruction of the boundary will significantly improve compared to the initial ARM reconstruction.

The resulting reconstruction is again segmented, resulting in a new partition of the image into interior, background, and boundary pixels. As new pixels can be added to the boundary, pixels whose values were fixed in the previous step can now become boundary pixels and vice versa. The procedure of alternating segmentation and reconstruction steps is then iterated until a pre-defined convergence criterion is reached.

A well-known limitation of DART is that the result can depend sensitively on the choice of gray levels and this choice may not be correct based on the first SIRT iterations. Sophisticated algorithms for choosing the correct gray levels can be found in \cite{graylevel1,graylevel2}.

\subsection{Object-based reconstruction methods}
The second category of reconstruction methods that we consider in this paper is object based in the sense that the routines aim to determine a small number of parameters that completely describe the object (in our case, the vertices of polytopes).

\subsubsection{Gardner-Kiderlen X-ray (GKXR)}\label{subsecGKXR}
The Gardner-Kiderlen X-ray algorithm \cite{GardnerKiderlen} is a recent development from the field of geometric tomography. It arose from theoretical work \cite{GarM80} in which it was shown that there are certain sets of four directions in 2D such that the exact projections of a 2D convex object in these directions determine it uniquely among all 2D convex shapes.  For example, directions specified by the four vectors $(0,1)$, $(1,0)$, $(2,1)$, and $(-1,2)$ constitute such a set \cite{GarG97}.  The GKXR algorithm is based on the simple observation that given a sufficiently dense set of lines meeting a convex set $K$, the convex hull of all the points at which the lines intersect the boundary of $K$
will form a convex polygon that approximates $K$ well.  The algorithm attempts to find this polygon for the set of projection measurement lines.

\begin{figure}[htb]
\begin{center}
\includegraphics[width=0.5\columnwidth]{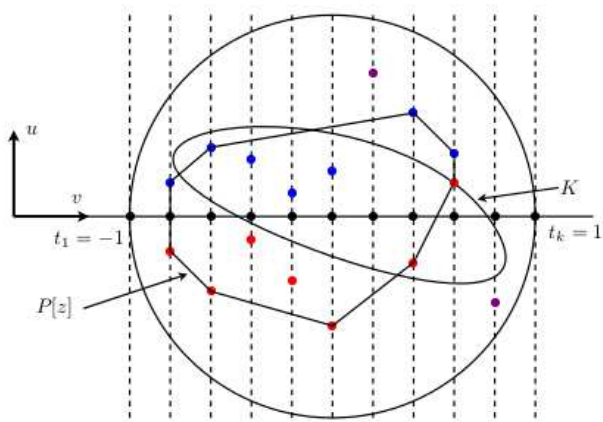}
\end{center}
\caption{Illustration showing the basic principles behind the GKXR
algorithm.} \label{fig:GKXR}
\end{figure}

Fig.~\ref{fig:GKXR} shows a schematic diagram of the basis of the algorithm.  The unknown object is the oval $K$, assumed to lie inside the circle.  For clarity, only a single projection, taken in the direction $u$, is considered in Fig.~\ref{fig:GKXR}, although in practice projections in four different directions are used.  For each projection direction $u$, detector pixels are located at the equally spaced points $t_1,\dots,t_k$ on the axis in the orthogonal direction $v$. The dotted lines through these points represent measurement lines.  A pair of points (in Fig.~\ref{fig:GKXR}, one red and one blue, shown in purple if they coincide) is placed randomly on each of the 4$k$ measurement lines.  Since the geometry of the measurement lines is known, the position of each point can be described by a single real variable giving the location of the point on the measurement line.  Therefore the position of all of the points can be described by a single vector variable $z$ with 8$k$ real components.

An initial guess for $K$ is obtained by forming the convex hull of all 8$k$ points, except those for which a pair coincides, i.e., the purple points.  This is the convex polygon labeled $P[z]$ in
Fig.~\ref{fig:GKXR}.  The convex hull is computed using a standard algorithm as a subroutine.  The reason for ignoring the purple points in taking the convex hull is that if a measurement line does not meet $K$, there must be some mechanism to eliminate the pair of points that lies on that line.  In practice, a threshold is set so that a pair of points is eliminated if they become too close in the iterative optimization procedure to be described next.

In order to improve the initial guess, the positions of the pairs of points on the measurement lines must be adjusted.  This is effected by computing the sum, over all measurement lines, of the squares of the differences between the measured projection value for $K$ and the corresponding projection value of
$P[z]$.  This least squares sum is the objective function in an optimization problem with 8$k$ real variables and an optimization routine is used to drive the value of the objective function down to a minimum.  The output of the algorithm is the convex polygon $P_k=P[z]$ corresponding to the optimal vector $z$ of these real variables.

In \cite{GardnerKiderlen} it is shown that for any finite set of directions for which the corresponding exact projections determine a convex object uniquely, the output $P_k$ converges to $K$ as $k\to \infty$, even when the measurements are affected by Gaussian noise of fixed variance.  Moreover, this remains true even if the optimization problem is not solved exactly, but only within an error $\varepsilon_k>0$, provided $\varepsilon_k\rightarrow 0$ as $k\to\infty$; see \cite[p.~337]{GardnerKiderlen}.  In practice, the optimization problem involved is heavily non-linear.  The {\tt fmincon} function from Matlab's Optimization Toolbox was used, along with simulated annealing to improve performance.

\subsubsection{Unfiltered backprojection (U-FBP)}
U-FBP, as described here, can be viewed as a geometric method. The idea is to backproject the object's shadows (yielding a strip for each shadow) and to return the intersection of all of these strips. The returned object is then necessarily a convex polygon. This, in general, cannot be guaranteed for other common U-FBP variants that backproject projections instead of shadows.

\subsubsection{Modified Prince-Willsky (MPW)}
The modified Prince-Willsky algorithm \cite{gardnerkiderlen09}, a modification of the algorithm in \cite{princewillsky90}, reconstructs a convex object $K$ from its \emph{support function} $h_K$. The function $h_K$ takes a direction (unit vector $u$) as input and returns a number that corresponds to the extent of $K$ in direction $u$. To be precise,
$$ h_K(u) = \max_{x \in K} u^T x,$$ where $u^Tx$ denotes the inner product of $u$ and $x$.

\begin{figure}[ht]
\centering
\includegraphics[width= 0.8\columnwidth]{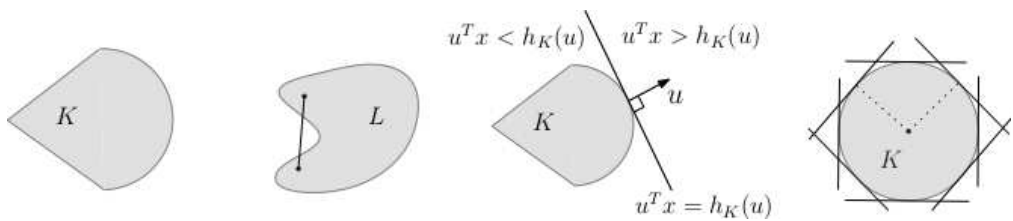}
\caption{Left to right: A convex set $K$ in the plane; a set $L$ which is not convex; the support function of $K$ in one direction $u$; support function values in many directions describing the convex set $K$. \label{fig:support}}
\end{figure}

Fig.~\ref{fig:support} also indicates that $K$ is completely determined by its support function values in all directions (this can be shown mathematically; see \cite[Section~0.6]{gardner}). Good approximations can already be obtained using a finite number of (suitably chosen) directions.

The support function values of $K$ in the two directions perpendicular to the projection direction can be easily determined from the data, because they correspond to the minimal (respectively, maximal) coordinates of the pixels in the data that record non-zero intensities. As data are available for many tilt angles, support function measurements are collected for different $u$ vectors.  These serve as input to the algorithm.

\begin{figure}[ht]
\centering
\includegraphics[width=0.8\columnwidth]{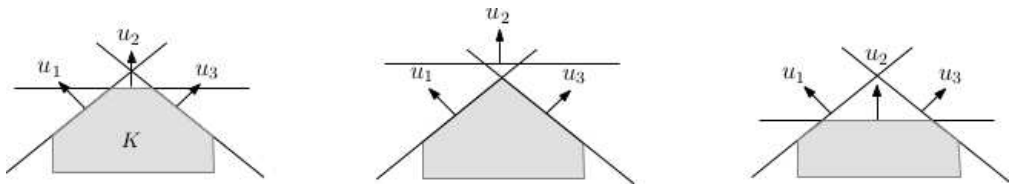}
\caption{Left: A convex set $K$ and three exact measurements of its support function. Middle: An incorrect support function measurement leading to inconsistent data. Right: An incorrect support function measurement cutting away a big part of $K$. \label{fig:supportNoise}}
\end{figure}

So far, the algorithm is very similar to U-FBP. However, U-FBP tries to find an object that fits the noisy measurements perfectly. As with most inverse problems, this is usually not the best strategy. Fig.~\ref{fig:supportNoise} illustrates the fact that noise may lead to inconsistencies in the data.

The MPW algorithm is designed to deal with noise. More precisely, for (Gaussian) noise affected measurements $h_1,\dots, h_n$ of the support function of $K$ in a finite number of directions $u_1,\dots,u_n$, the MPW algorithm solves a (linearly) constrained least-squares problem to obtain values $y_1, \dots , y_n$, which are the support function values of the best-approximating set $K^*$.

With mild restrictions, the output of the algorithm converges as the number of shadows, affected by Gaussian noise of fixed variance, approaches infinity \cite{convergence}. The implementation of MPW is somewhat more demanding than U-FBP, because a subroutine for solving quadratic programs is required (we use the Xpress solver). The set $K^*$, in our implementation, is obtained as an intersection of halfspaces $K^*=\bigcap_{i=1}^n\{x \: : \: u_i^Tx \leq y_i\}$ via U-FBP. Additional MPW variants are discussed in \cite{gardnerkiderlen09}.

\subsubsection{2n-GON}
Again, shadows are taken as input data. We aim to reconstruct an
object $K$ that is known to be close to a regular $2n$-gon, where $n
\geq 3$ is known in advance. (In our experimental application, $n=3$.) If
the assumptions are not fulfilled then the algorithm should exit without
reconstruction.

The length of the shadow of $K$ for a given tilt angle
$\theta$ is commonly referred to as the \emph{width of $K$ orthogonal to
$\theta$}. Of course, this quantity can easily be computed from the
input data. It is easy to see that if $K$ is a regular $2n$-gon, then
the width of $K$, as a function of $\theta \in [0^\circ,180^\circ)$, has
exactly $n$ local minima, corresponding to the tilt angles that project
$K$ along an edge direction of $K$. These $n$ minima are thus
$(180/n)^\circ$ apart and if they can be determined, then unfiltered
backprojection (as implemented in U-FBP) from the corresponding $n$
directions yields the regular $2n$-gon $K$. Note that the shadows from
two such edge directions of a regular $2n$-gon $K$ determine
the minimum width and center of $K$ and hence $K$ itself. Also note
that two such shadows are typically available from the data, because
standard electron microscopes allow  $[0^\circ,120^\circ)$ tilt ranges.

\begin{figure}[ht]
\centering
\includegraphics[width=0.6\columnwidth]{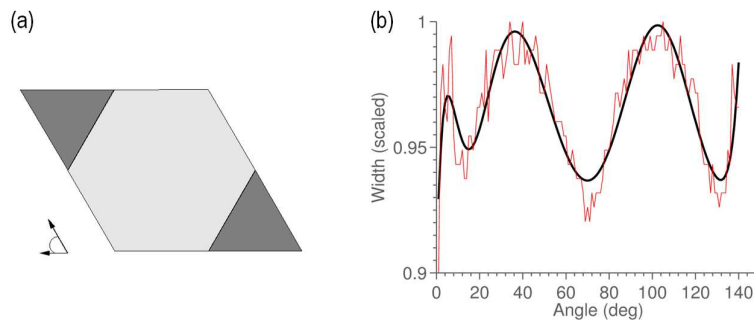}
\caption{(a) A regular hexagon (lightly shaded) and the parallelogram determined by width data from angles indicated by the arrows; (b) measured width as obtained from the HAADF STEM data (red) and best fitting (in the least-squares sense) polynomial curve of degree $11$ (black).}\label{fig:hexagon}
\end{figure}

The procedure also works when $K$ is only close to a regular $2n$-gon,
as follows. Again, the idea is to apply unfiltered backprojection for
the directions orthogonal to the tilt angles determined by some $n$
minima of the width function of $K$ that are only approximately $(180/n)^\circ$ apart.  As before, some of the data for edge projecting directions might not be available if the data are not acquired over the full $[0^\circ,180^\circ)$ tilt range. If this is the case, then one needs to impose some assumptions on the missing data; see Fig.~\ref{fig:hexagon}(a), in which the hexagon and the parallelogram are indistinguishable from data in the (very limited) $[0^\circ,70^\circ)$ tilt range shown. Another limitation of the $2n$-GON approach is that we need to assume that the noise level in the (shadow) data is sufficiently low to allow determination of the minima of the width function.

Here is a precise description of our implementation. We assume that data are available over a $[0^\circ,\omega^\circ)$ tilt range, where $\omega$ is fixed (typically $\omega \approx 140$). We initially determine the local minima in $[0^\circ,\omega^\circ)$ of a polynomial curve of degree at least $k \geq 2n$ that best fits (in the least-squares sense) the measured widths. We found in simulations that values of $k$ around $2n+5$ give good and stable results. Fig.~\ref{fig:hexagon}(b) shows measured widths of a hexagon together with its best fitting polynomial curve.

Suppose there are $m\geq 2$ minima $t_1 \leq t_2 \leq \cdots \leq t_m$, in degrees, within the $[0^\circ,\omega^\circ)$ range (otherwise, we stop without reconstruction).  Unfiltered backprojection from $S=\{t_1,\dots,t_m\}$ yields a (possibly degenerate) $2m$-gon, which, for $\omega=180$, approximates $K$.  If $\omega<180$, we proceed as follows.  Let
$$
R=\left\{i \in \{1,\dots,m-1\} \: : \: 180/n-10 \le |t_{i+1}-t_i| \le 180/n+10\right\}.
$$
If $R=\emptyset$, we stop without reconstruction.  If $|R|=n-1$, we reconstruct from the angles in $R$.  Otherwise, $1\le |R|=r<n-1$ and we define $a=\max\{R\}$, $b=a+1$, and $T=\{t'_1,\dots,t'_{n-r}\}$, where
$t_i'=t_b+180i/n$, $i=1,\dots,n-r$.   If there is an angle in $T$ that is not in the $[0^\circ,180^\circ)$ range, we exit without reconstruction. Otherwise, we reconstruct from $S\cup T$.  Here if $t_i'$ is outside the tilt range, we use unfiltered backprojection of the shadows for the angles $t_a$ and $t_b$ to obtain a parallelogram and set the shadow for $t_i'$ to be equal in length to that for $t_b$, using the center of the parallelogram to position it correctly.

We remark that this implementation reduces in the regular $2n$-gon case to the method described at the beginning of this subsection.
Note that $2n$-GON can also be applied to data from a small number of tilt angles, since the best fitting polynomial curve may still approximate the local minima, although these minima might not be present in the data. We test this, among other things, in the next two sections. A theoretical analysis, however, remains outside the scope of this paper, as tolerable deviations from a regular $2n$-gon depend on several parameters such as the relative position of the missing wedge, the amount of noise in the data, the number of available projections, and the diameter of the $2n$-gon.

It is worth mentioning that as well as measuring widths orthogonal to $\theta$, we can also measure widths parallel to $\theta$, provided that projection data (and not only shadows) are available. We shall not discuss the performance of this variant here.

\section{Experimental application} \label{ex:section}
We consider the task of reconstructing a nanowire from HAADF STEM data. Nanowires, small wires that are tens of nanometers in diameter and micrometers in length, are promising building blocks for future electronic and optical devices; see \cite{nanowireappl1, nanowireappl2}. They are typically grown from a substrate and much research effort is being focused on understanding and controlling their growth mechanisms \cite{nanogrowth}. Electron tomography, as in various materials science applications, is rapidly developing into a powerful 3D imaging tool for studying these effects at the nanoscale \cite{tomobookmater, electrontomoreview}.

With current technology, the tomographic data acquisition time for 140 projections of a single nanowire is about 2 hours when performed manually. This is currently a bottleneck preventing many in-situ experiments on short time scales and the imaging of multiple nanowires. Automated acquisition can lower this time, but a further reduction, which could be achieved if the required number of tilt angles can be reduced, is of paramount importance. Computation times for the reconstruction of a single $512\times 512$ slice range from a few seconds to $3$~minutes (for GKXR) on a standard PC.

The particular nanowire in our experimental application is grown from pure InAs. Such nanowires are usually convex, or nearly so, in cross-sections perpendicular to the growth direction. In fact, most of these cross-sections for InAs nanowires are close to regular hexagons \cite{hexagoncrosssection}.

\subsection{Experiment}
An HAADF STEM image series of an InAs nanowire was acquired using a probe-aberration-corrected FEI Titan $80$--$300$ microscope operated at $300$ kV with a probe convergence semi-angle of $18$ mrad and an inner detector semi-angle of $100$ mrad. The images were acquired over a total angular range of $139^\circ$  with a $1^\circ$ tilt increment. The original $2048\times 2048$ pixel images, representing the projections for $2048$ slices, were then binned by a factor of $4$ to reduce noise in the reconstruction. The resolution after binning corresponds to a pixel size of $0.84$ nm $\times$ $0.84$ nm.

\begin{figure}[hbt]
\begin{center}
\resizebox{0.8\columnwidth}{!}{\includegraphics{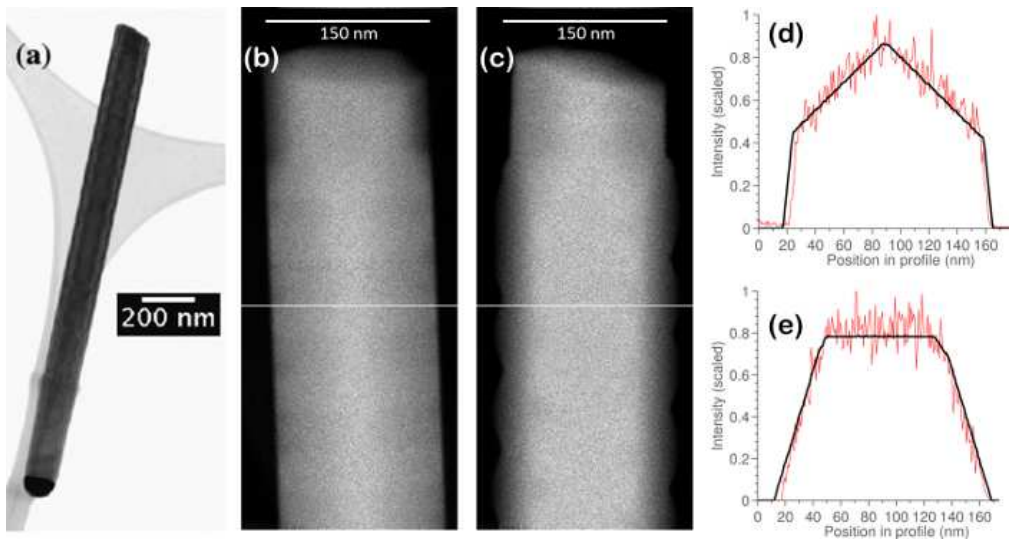}}
\end{center}
\caption{Nanowire data: (a) bright-field TEM image of the InAs nanowire specimen used for tomography; (b,c) aligned binned HAADF STEM images taken from the tilt series of the wire at angles of $10^\circ$ and $40^\circ$, respectively (slice $220$ is indicated by the white line; the tilt axis is in the vertical direction through the center of the image); (d,e) measured projections of slice $220$
(non-linear projection intensities) shown in red and ideal projections of slice $220$
(linear projection intensities) shown in black, at angles of $10^\circ$ and $40^\circ$, respectively. (The ideal projections were estimated from our reconstructions.)}\label{fig1}
\end{figure}

Fig.~\ref{fig1}(a--c) show a bright-field transmission electron microscopy (TEM) image and two HAADF STEM images, respectively, of the InAs nanowire specimen. An effect of non-linear projection intensities within a particular slice is visible in Fig.~\ref{fig1}(d,e), as the measured projections (red graphs) do not exactly match the projections of the best-fitting shapes of the nanowire for this slice (black graphs). Twinning in the growth direction produces the ``bee stripe'' patterns visible in Fig.~\ref{fig1}(b,c).

\subsection{Parameters for the algorithms} \label{specifics1}
The relaxation parameter $\lambda$ for SIRT and DART was set to $1$. The SIRT algorithm was run for 50 iterations, based directly on the measured intensity values in the projection data. Afterwards, the reconstruction was thresholded at a value of $0.5$, where $0$ represents the background and $1$ represents the gray level of the nanowire. Each run of the DART algorithm consisted of $25$ SIRT iterations to compute the starting solution, followed by $25$ DART iterations, each of which included $10$ SIRT iterations on the set of free pixels (i.e., the pixels near the boundary). The SIRT iterations within each DART iteration only take a fraction of the time of a complete SIRT iteration, as they are only applied to the free pixels. As for SIRT, DART employs a threshold at $0.5$. A fixed fraction of $0.85$ was used, meaning that $85\%$ of all non-boundary pixels (selected randomly) are kept fixed in each DART step; the fixed fraction can be adapted to a specific noise level for more accurate results. We refer to \cite{DART2} for details.

The BART parameters, along with the code, are provided in Appendix~A, Table~\ref{table:BART}.

For reconstructions with GKXR, the projection in each direction was measured at 40 equally spaced positions along an axis orthogonal to the direction and passing through the center of a circular window known to contain the object to be reconstructed (i.e., $k=40$ in Fig.~\ref{fig:GKXR}).  Simulated annealing was used, with a fixed cooling schedule.  The projection data was pre-processed by a simple smoothing algorithm from digital signal processing, an averaging filter based on the IIR (infinite impulse response) design (see, for example, \cite[Chapter~8]{Lyo01}).  More specifically, if the non-zero 512 projection measurements for a certain direction are $y_1,\dots,y_m$, a smoothed set of measurements $y_1',\dots,y_m'$ is produced by recursively defining
$$y_i'= (y_i+y'_{i-1})/2,$$
for $i=1,\dots,m$, where we set $y_0=0$.  This smoothing procedure was iterated 50 times to obtain the smoothed data for input into the algorithm.

The algorithms U-FBP, MPW, and $2n$-GON require shadows as input. Let $M_\theta$ denote the $512 \times 512$ matrix containing (row-wise) the projections of the nanowire slices for viewing angle $\theta$. For this particular data set, we binarize the median filtered $M_\theta$ (with a $1\times 3$ window) using threshold $T=0$, which is followed by a morphological opening  \cite[Chapter~15]{klette} with Matlab's structuring element \texttt{strel('diamond', 2)}. This morphological opening takes neighboring slices into account; the reconstruction of the object, however, proceeds slice by slice.

\subsection{Experimental results}

\begin{figure}[ht]
\centering
\includegraphics[width=1\columnwidth]{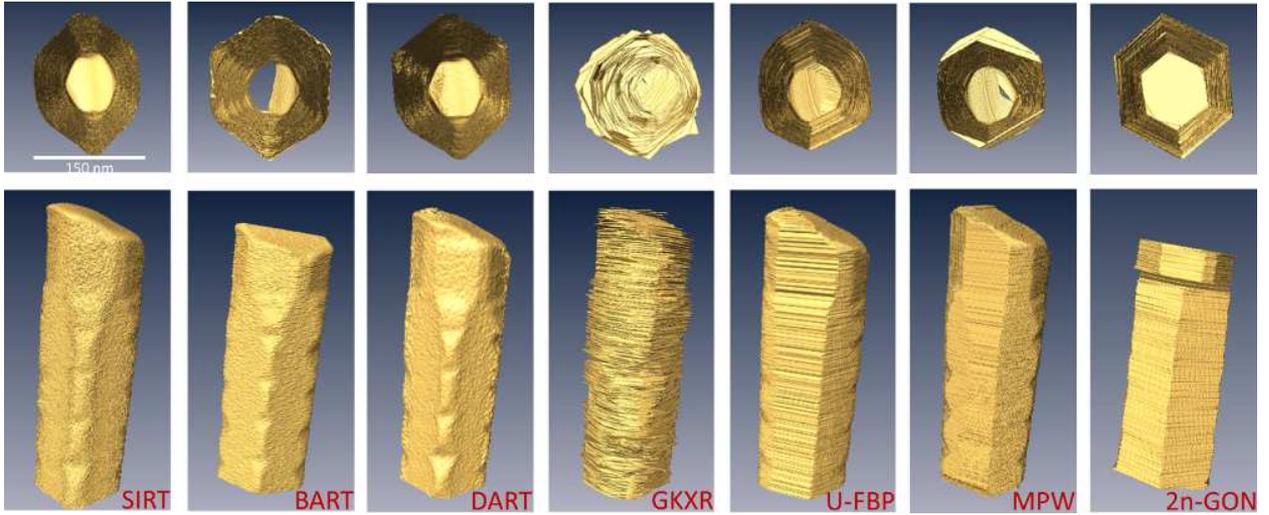}
\caption{Reconstruction of the nanowire using different algorithms. Top-to-bottom and frontal views are shown in the first and second row, respectively. GKXR requires only $4$ projections; U-FBP, MPW, and $2n$-GON reconstruct from shadows. A few of the top slices of the BART reconstruction are missing, because we put no effort into recovering the correct factor that relates the line integrals and the actual measured intensities.  Slices are missing in the $2n$-GON reconstruction if the algorithm detects that there is no hexagon in the corresponding slice. Nanowire orientations and their top-to-bottom views might vary as the 3D viewing points for the individual images have been manually selected. The $30$ degree rotation of the end of the wire relative to the main part is more readily discernible in the frontal views (lower row) than in the top-to-bottom views (upper row). Images have been rendered using the Amira software with the isosurface operation and the \texttt{compactify} option, but no explicit smoothing was applied. \label{fig:realdata}}
\end{figure}

A full discussion of our results is only possible in connection with the simulations presented in the next section, particularly because there are no data for this nanowire obtained by an independent imaging method. However, we can make some general remarks.

\begin{enumerate}[(i)]
\item It appears that the nanowire consists of two parts, the top part slightly narrower and rotated by about 30$^{\circ}$ around the axis of the bottom part. The bottom part seems to consist of alternating twins, each around $40$ nm thick, while one twin orientation dominates the top $80$ nm.  Except for the twinning (much less visible in GKXR and $2n$-GON), these structures can be found in each of the seven reconstructions. However, the region around the frontal vertical edge of the reconstructions lies in the missing wedge and so might contain artefacts. Perhaps the most reliable indication of the presence of twinning is the fine periodic structure that appears along the left- and right-most vertical edges of the reconstructions. This periodic structure is too fine to be resolved by our current implementation of GKXR.
\item BART and DART, followed by SIRT, appear to give the best (most accurate and precise) results. Note, however, that the software (Amira) that renders the 3D images in Fig.~\ref{fig:realdata} smooths out isolated pixels due to the surface mesh that is built into the isosurface operation.  Therefore these images might look smoother than those produced by these three algorithms without this post-processing.  This is not the case for the other four algorithms, because they do not return objects containing isolated pixels.  An explanation for the fuzzy boundary returned by GKXR is given in Section~\ref{synthsect:disc}.
\item It seems possible to infer useful geometric parameters about the facet structure (diameters, angles, etc.) from GKXR, U-FBP, MPW, and $2n$-GON, which use fewer data (shadows or fewer projections, as appropriate). See also Section~\ref{synthsection:disc3}.
\item Algorithms U-FBP, MPW, and $2n$-GON need segmentation of the shadows, while SIRT, BART, and DART need segmentation of the reconstruction. Depending on the reliability of the measured intensities, one type of segmentation (possibly involving filtering and thresholding) might be favorable over another.
\item It should be possible to improve the performance of GKXR by replacing the pre-processing via IIR with a least-squares fit to the projection data of a piecewise linear concave curve. Further improvement of the reconstruction quality of each of the presented algorithms might be achieved by incorporating the fact that the slices of the object are not independent from each other. The reconstruction from GKXR, for instance, might benefit from post-processing, such as a simple averaging process to smooth out the differences between consecutive slices.
\end{enumerate}

\section{Computer simulations} \label{synthsect}
\begin{figure}[ht]
\centering
\includegraphics[width= 1.0\columnwidth]{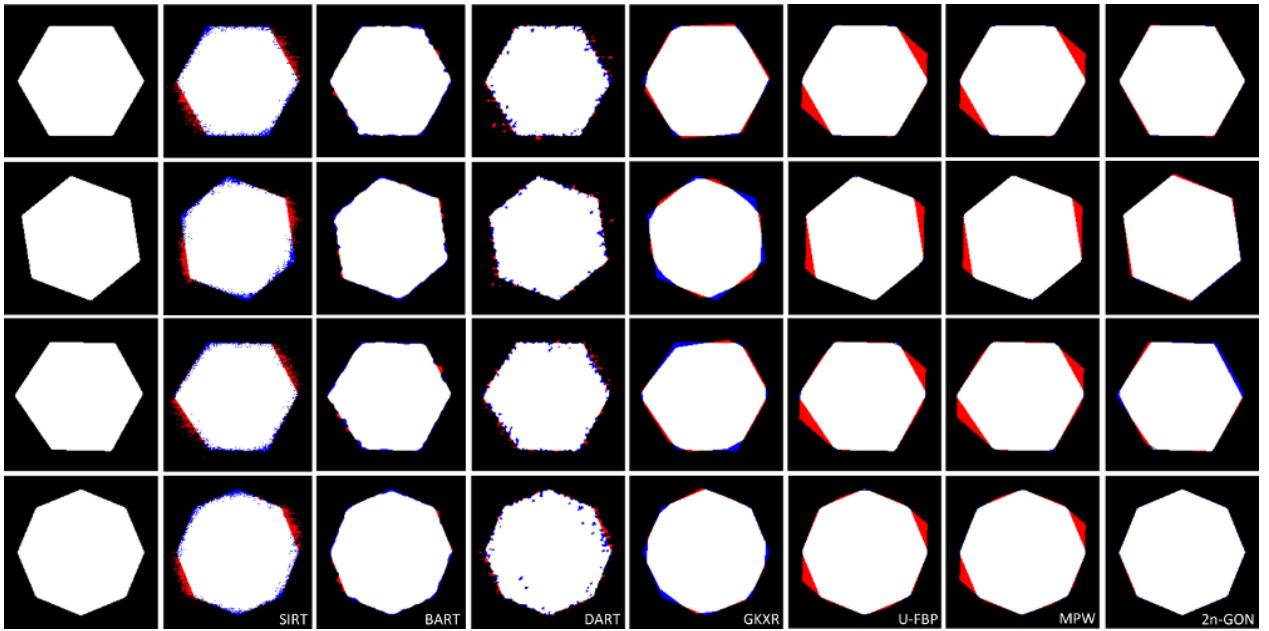}
\caption{Phantoms and reconstructions~I. Column~1: Phantoms 1--4 (from top to bottom).
Subsequent columns show difference images between a phantom and a typical reconstruction with $S_{140,10}$ at $50$-noise obtained by SIRT (Column~2), BART (Column~3), DART (Column~4), GKXR (Column~5), U-FBP (Column~6), MPW (Column~7), and $2n$-GON (Column~8).  Color scheme for difference images: White  pixels belong to the phantom (P) and the reconstruction (R), red pixels to $R\setminus P$, and blue  pixels to $P\setminus R$. These are $192\times192$ pixel images in which the $160$ pixel boundary has been cropped. \label{fig:fig7}}
\end{figure}

We tested the algorithms with four phantoms (i.e., simulated objects) under varying magnitudes of noise. As phantoms we used two regular hexagons, a slightly irregular hexagon, and a regular octagon. They are shown in Fig.~\ref{fig:fig7} and henceforth are referred to as Phantoms 1--4.

To compare phantoms and reconstructions we need to quantify the degree to which two shapes differ. (This is a central problem in computer vision.) Here we use two metrics that are frequently encountered in the literature, the \emph{symmetric difference metric} and the \emph{Hausdorff metric}. The \emph{symmetric difference distance} between finite sets $A$ and $B$ of pixels (in our case, subsets of $\{1,\dots,512\}\times\{1,\dots,512\}$) is defined by
$$
\delta_S(A,B)=\left|(A\setminus B)\cup (B\setminus A)\right|,
$$
i.e., $\delta_S(A,B)$ counts the number of mismatching pixels. The corresponding \emph{Hausdorff distance} is defined by
$$
\delta_H(A,B)=\max\left(\delta(A,B),\delta(B,A)\right),
$$
where
$$
\delta(A,B) = \max_{a \in A}\:\:\min_{b \in B} ||a-b||_\infty= \max_{(a_1,a_2) \in A}\:\:\min_{(b_1,b_2) \in B}\:\: \max\left(|a_1-b_1|,|a_2-b_2|\right)
$$ and $|\cdot|$ denotes the usual absolute value of a real number. (We chose the $L_\infty$ norm for computational convenience; other choices are possible.) In other words, for $\delta(A,B)$ we identify the pixel $a \in A$ that is farthest from any pixel in $B$ and return the distance from $a$ to its nearest neighbor in $B$. Taking the maximum of $\delta(A,B)$ and $\delta(B,A)$ makes $\delta_H$ symmetric in its arguments. Thus, roughly speaking, $\delta_H(A,B)$ measures the extent to which each pixel of $A$ lies near some pixel of $B$ and vice versa.

Now, let $P$ denote the set of pixels of a phantom and $R$ the set of pixels of a reconstruction. We refer to $\delta_S(P,R)$ and $\delta_H(P,R)$, or $\delta_S$ and $\delta_H$ for short, as \emph{reconstruction errors} (measured in the symmetric difference metric and Hausdorff metric, respectively).

The Hausdorff and symmetric difference metrics have different and somewhat complementary characters. While the Hausdorff metric is sensitive to single pixel outliers but robust to boundary perturbations, the symmetric difference metric is robust to single pixel outliers but sensitive to boundary perturbations (cf.~Fig.~\ref{fig:fig7}, particularly the DART reconstructions). However, as for any metric, two sets are equal if and only if they have zero distance.

\subsection{Data generation} \label{sect:datageneration}
To avoid the ``inverse crime'' \cite[Section~5.3]{coltonkress} of using the same model for generating the data for testing the algorithms as the one used in their design, we generate the projections from higher resolution versions of the phantoms (cf. \cite{kaipio, alpersphase}). More specifically, from $2048\times 2048$ pixel versions of the phantoms we generate the projections and bin them by a factor of $4$. We thus aim at reconstructing the $512\times 512$ pixel versions of the phantoms as shown in Fig.~\ref{fig:fig7}. No further scaling of the projection values is introduced, i.e., the projection values give the number of pixels of the phantom that lie on the corresponding line. (Our implementation employs the Matlab command \texttt{imrotate} using the \texttt{bilinear} and \texttt{crop} option.) Comparing our results to those obtained from projection data generated by SNARK09 \cite{snark09}, we could not find significant differences. This could be expected, because our projections (before binning) are generated from suitably high resolution phantoms. The noise model that is described next is applied to the projections generated from the high resolution phantoms.

Taking a simplistic approach, we simulate Gaussian noise. Hence, we specify noise by one parameter $\sigma$. We draw independent and identically distributed zero-mean, $\sigma^2$-variance normal random variables $p(i,j)$ for each projection angle $i$ and projection pixel $j$ that has non-zero intensity. The $p(i,j)$'s are added to the intensities of the corresponding projection pixels; negative intensities are set to zero. This approach simulates additive Gaussian noise on the non-zero intensities.
The underlying assumption in our noise model of adding noise only to pixels with non-negative intensities is that statistical noise effects affect the recorded signal but still allow detection of the object's shadows. At least for our HAADF STEM data of the nanowire, this seems to be a plausible model. In our simulations we consider the $0$-noise (i.e., $\sigma=0$, noise-free) and $50$-noise ($\sigma=50$) cases. Here $50$-noise seems to be in agreement with the intensity variations present in the experimental data in Section~\ref{ex:section}.

\subsection{Parameters for the algorithms} \label{specifics2}
For our algorithms, the parameters used for all simulations were the same as for the nanowire slice reconstruction described in Section~\ref{specifics1}. The pre-processing of the data was simplified in the sense that for GKXR the pre-processing procedure to smooth the data was not used in the $0$-noise case and for U-FBP, MPW, and $2n$-GON the input shadows were obtained directly by thresholding the projections with $T=0$.

\subsection{Simulation results} \label{synthsect:disc}
\begin{figure}[htb]
\includegraphics[width=1.0\columnwidth]{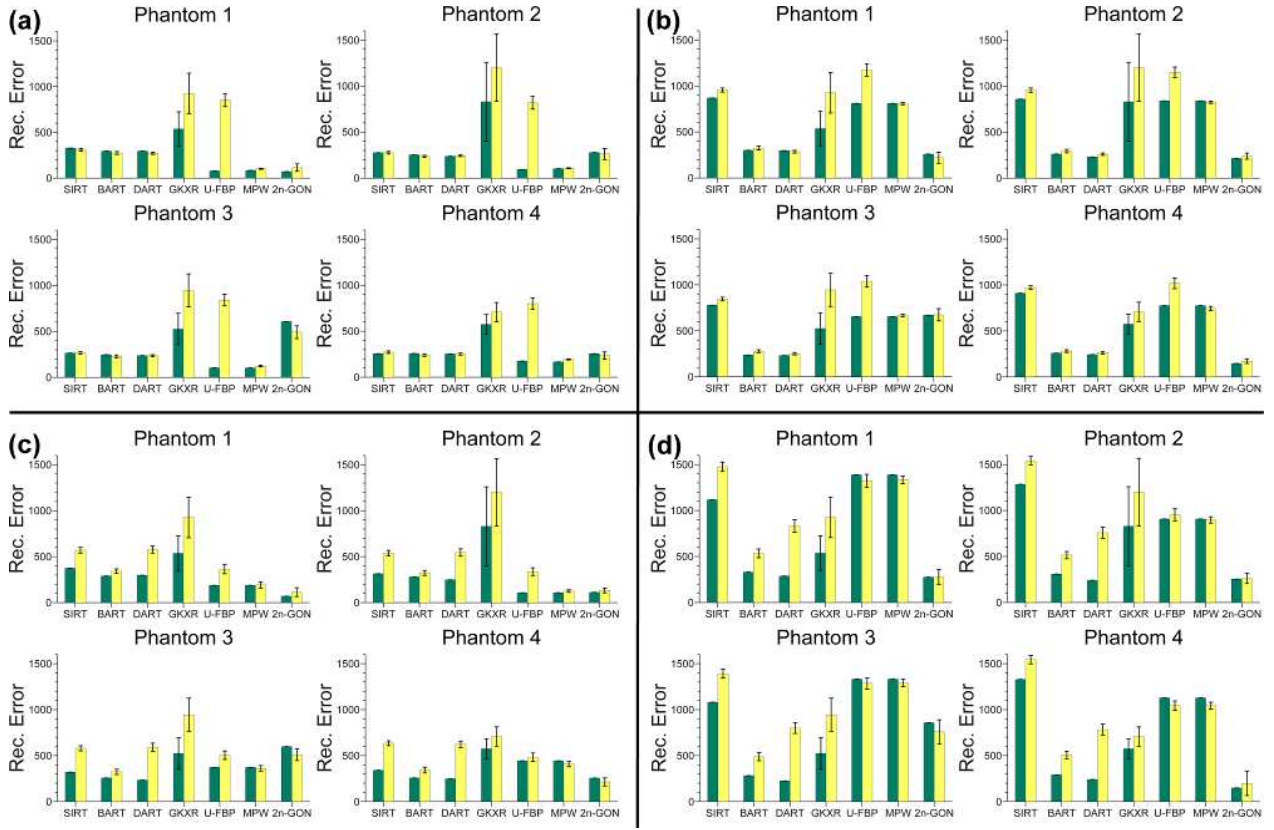}
\caption{Simulation results presented as mean reconstruction errors measured in the symmetric difference metric: (a) $S_{180,1}$, (b) $S_{140,1}$, (c) $S_{180,10}$, (d) $S_{140,10}$. Bar colors indicate noise level,
green for $0$- and yellow for $50$-noise. Black error bars represent standard deviation. GKXR requires only $4$ projections; U-FBP, MPW, and $2n$-GON reconstruct from shadows.}\label{fig:result2}
\end{figure}

\begin{figure}[htb]
\includegraphics[width=1.0\columnwidth]{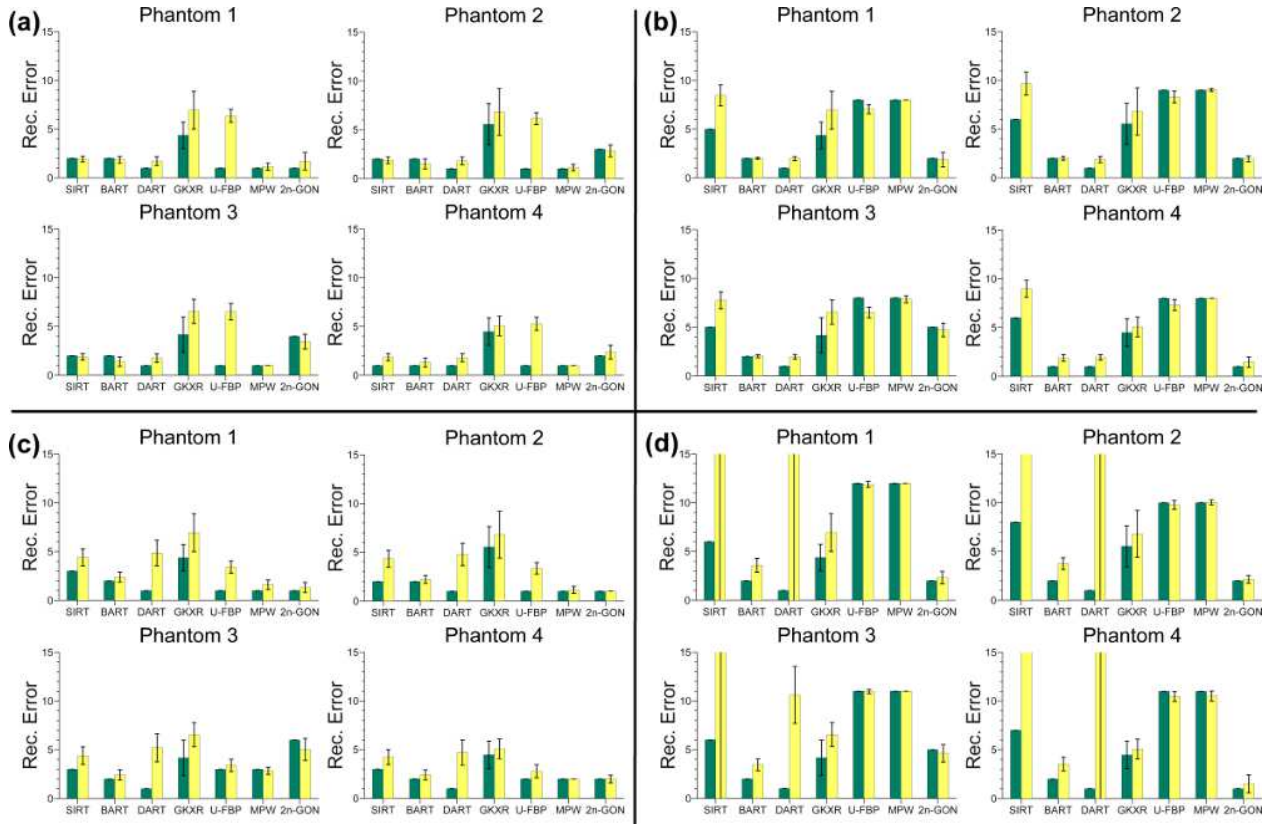}
\caption{Simulation results presented as mean reconstruction errors measured in the Hausdorff metric: (a) $S_{180,1}$, (b) $S_{140,1}$, (c) $S_{180,10}$, (d) $S_{140,10}$. Bar colors indicate noise level,
green for $0$- and yellow for $50$-noise. Black error bars represent standard deviation. GKXR requires only $4$ projections; U-FBP, MPW, and $2n$-GON reconstruct from shadows. (For clarity, we do not show the bars of the SIRT and DART results for $S_{140,10}$ with $50$-noise if they extend beyond the value $15$; their actual values range between $20$ and $160$.)}\label{fig:result1}
\end{figure}

We employed the following four sets of tilt angles: $S_{180,1}=\{1^\circ,2^\circ,3^\circ,\dots,180^\circ\}$,
$S_{140,1}=\{1^\circ,2^\circ,3^\circ,\dots,140^\circ\}$, $S_{180,10}=\{1^\circ,11^\circ,21^\circ,\dots,171^\circ\}$, and $S_{140,10}=\{1^\circ,11^\circ,21^\circ,\dots,131^\circ\}$. Here a tilt angle $\theta$ corresponds to a clockwise rotation of angle $\theta$ around the vertical $(0,1)$-direction. While $S_{140,1}$ corresponds to our experimental setting, we included the other sets of tilt angles to compare the performances of the algorithms with respect to the missing wedge and the total number of available tilt angles. For the main simulations reported in Figs.~\ref{fig:result2} and~\ref{fig:result1}, reconstructions with GKXR were always performed using the four angles $\{1^\circ, 28^\circ, 91^\circ, 118^\circ\}$, chosen to be nearest to a theoretically ideal set within the range $1^\circ$--$140^\circ$, namely angles corresponding to the vectors $(0,1)$, $(1,2)$, $(1,0)$, and $(2,-1)$ (cf.~Section~\ref{subsecGKXR}).

For each algorithm, each set of tilt angles, and each noise level we performed $100$ reconstructions. The mean reconstruction errors $\delta_S$ and $\delta_H$ together with their standard deviations are shown in Fig.~\ref{fig:result2} and~\ref{fig:result1}, respectively.

We first discuss typical difference images of reconstructions of each phantom, shown in Columns~2--8 of Fig.~\ref{fig:fig7}, obtained from SIRT, BART, DART, GKXR, U-FBP, MPW, and $2n$-GON with tilt angles $S_{140,10}$ at $50$-noise. Note that this means reconstruction from only $14$ projections, ten times fewer than is typically employed in ET (the $S_{140,1}$ case). White pixels in the color scheme of our difference images correspond to correctly reconstructed pixels, red pixels belong to the reconstruction but not the phantom, and blue pixels belong to the phantom but have not been reconstructed. The reconstruction errors ($\delta_S$; $\delta_H$) for Phantoms~1--4, respectively, are:

\begin{center}
\begin{tabular}{l l l l l}
SIRT: & $(1,482;13)$ & $(1,562;15)$ & $(1,367;13)$ & $(1,594;16)$;  \\
BART: & $(523;3)$ & $(513;3)$ & $(491;4)$ & $(494;4)$; \\
DART: & $(873;16)$ & $(697;14)$ & $(843;7)$ & $(955;8)$; \\
GKXR: & $(693;4)$ & $(1,174;7)$ & $(933;6)$ & $(670;5)$; \\
U-FBP: & $(1,285;12)$ & $(922;10)$ & $(1,258;11)$ & $(1,022;10)$; \\
MPW: & $(1,282;12)$ & $(905;10)$ & $(1,227;11)$ & $(948;10)$; \\
$2n$-GON: & $(215;2)$ & $(386;3)$ & $(653;4)$ & $(146;1)$.
\end{tabular}
\end{center}

\smallskip

\noindent The difference images illustrate several characteristics of the different algorithms. For instance, the algorithms that are specifically designed to reconstruct from only a few directions, GKXR and $2n$-GON, yield small errors, both in the symmetric difference metric and the Hausdorff metric.  (Recall that GKXR uses only four projections.) On the other hand, the effect of the missing wedge can be seen in the $2n$-GON and MPW reconstructions, as these algorithms reconstruct from shadows. The $2n$-GON algorithm cannot fully compensate for the missing wedge, because Phantom~3 deviates from a regular structure.  The fuzziness of the object boundary in the BART and, particularly, the DART reconstruction is a common phenomenon for pixel-based reconstruction methods that do not employ filters.

\begin{figure}[ht]
\centering
\includegraphics[width= 0.8\columnwidth]{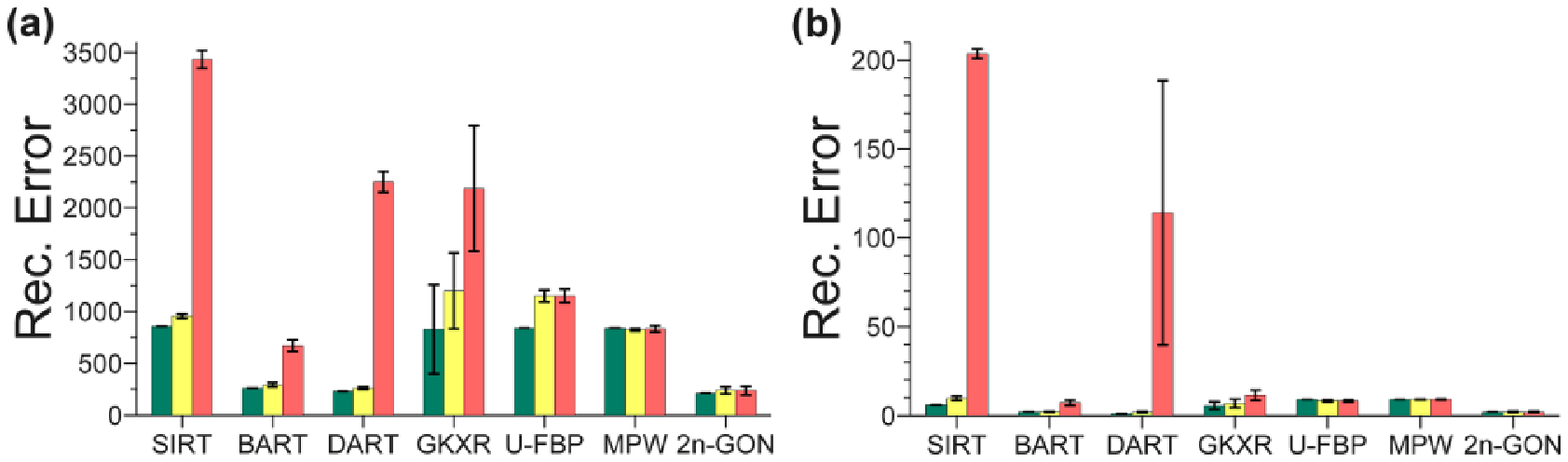}
\caption{Simulation results for Phantom 2 with $S_{140,1}$ including $200$-noise, presented as mean reconstruction errors measured in (a) the symmetric difference metric and (b)~the Hausdorff metric. Bar colors indicate noise level, green for $0$-, yellow for $50$-, and red for $200$-noise. Black error bars represent standard deviation. GKXR requires only $4$ projections; U-FBP, MPW, and $2n$-GON reconstruct from shadows.\label{fig:highnoise}}
\end{figure}

We now turn to a discussion of Figs.~\ref{fig:result2}--\ref{fig:highnoise}. We refer to $0$-noise and $50$-noise as \emph{moderate noise levels}, while $x$-noise with $x \geq 200$ is a \emph{high noise level}. At high noise levels, the reconstruction quality, except perhaps for MPW and $2n$-GON, becomes very poor. Therefore Figs.~\ref{fig:result2} and~\ref{fig:result1} only display results for moderate noise levels. Typical results for high noise levels are presented in Fig.~\ref{fig:highnoise} and discussed in~(\ref{highnoiseitem}) below.

The simulations for $S_{140,1}$ at $50$-noise resemble the experimental conditions of the nanowire reconstruction presented in the previous section; the corresponding simulation results are depicted by the yellow bars in Figs.~\ref{fig:result2}(b) and~\ref{fig:result1}(b). Experimental conditions with a reduced number of tilt angles would be represented by $S_{140,10}$ at $50$-noise. The key inferences from the simulations can be summarized as follows.

\begin{enumerate}[(i)]
\item For available projections over the whole $180^\circ$ tilt range (as in the $S_{180,1}$ and $S_{180,10}$-case) and moderate noise levels, we find that all algorithms yield good (accurate and precise) results (usually with $\delta_H\leq 5$ and $\delta_S \leq 1000$).
\item BART, DART, and $2n$-GON give the best results in the $S_{140,1}$ case at moderate noise levels (the $50$-noise case resembles the experimental setup).
\item \label{prob:2ngon} $2n$-GON performs well, even with missing wedge and few available projection directions. However, the object needs to be close to a regular $2n$-gon, otherwise caution is required (see the results for Phantom~3).
\item In many cases, BART and DART give the best results.  Particularly with DART, the quality of the reconstruction deteriorates significantly if fewer projections are available and if the noise level increases.
\item \label{prob:GKXR} According to the simulations, the GKXR algorithm is among the better-performing algorithms in the missing wedge case at moderate noise (see Figs.~\ref{fig:result2}(d) and~\ref{fig:result1}(d)) and its outstanding feature is that it requires projection data only from four directions. To test the effect of changing the four directions used, we performed the same simulations with angles $\{21^\circ, 41^\circ, 61^\circ, 81^\circ\}$, all contained in a narrow $60^\circ$ range.  The results were worse, but not dramatically so.  For example, for Phantom~1 at $50$-noise, the mean $(\delta_S;\delta_H)$ errors rose to $(1,227.83; 8.48)$, compared to $(925.84; 6.94)$ for angles $\{1^\circ, 28^\circ, 91^\circ, 118^\circ\}$.  However, the nanowire reconstruction in Fig.~\ref{fig:realdata} seems the worst produced by the algorithms used.  To some extent, this is due to the fuzzy nature of the boundary depicted there.  In fact, while the mean and standard deviations of the errors reported for GKXR in Figs.~\ref{fig:result2}(b) and~\ref{fig:result1}(b) are fairly small, the distance between neighboring slices of the nanowire reconstruction can be significant.  This is caused by the inherent stochastic nature of the algorithm, which uses simulated annealing.  Possible improvements have been discussed in Section~\ref{specifics1}.
\item The problems for MPW and U-FBP are caused by the missing wedge, because data are simply missing from the shadows (see Fig.~\ref{fig:hexagon}(a)). If essential object features do not lie in the missing wedge, then good reconstruction results can be obtained even at high noise levels.  In fact, MPW is among the better-performing algorithms if data are acquired over a $180^\circ$ angular range. The U-FBP algorithm never outperforms MPW by a significant margin and in most cases the results for MPW are clearly superior to those reported for U-FBP. Inspection of the reconstructions shows that U-FBP tends to cut off the corners of the object, while MPW returns the correct shape up to a systematic overestimation. See also (\ref{highnoiseitem}).
\item SIRT is known for its noise suppressing features. It performs much better in highly limited data scenarios than transformation based techniques such as \emph{filtered backprojection} \cite[Chapter~7]{kakslaney}. These findings are confirmed in our simulations. To avoid overloading the figures, however, we chose not to present the corresponding poorer results for those scenarios that we obtained by filtered backprojection. While SIRT gives rather good results as measured by the Hausdorff metric for $180^\circ$ angular range ($S_{180,1}$, $S_{180,10}$) at moderate noise levels, we observe from Figs.~\ref{fig:result2} and~\ref{fig:result1} that the algorithm is typically outperformed by the other algorithms in the missing wedge case ($S_{140,1}$, $S_{140,10}$).
\item \label{highnoiseitem} The reconstruction quality of all of the algorithms deteriorates at high noise levels. Fig.~\ref{fig:highnoise} indicates that U-FBP, MPW, and $2n$-GON seem to be less affected. (We show only results for Phantom~$2$ and $S_{140,1}$; the other cases are similar.) This finding is somewhat expected, since these algorithms work with shadow data.
\item There is a notable discrepancy between the performance of $2n$-GON  and GKXR in the $S_{140,1}$ simulations with $50$-noise and their performance, much worse in the case of GKXR, in the nanowire reconstructions shown in Fig.~\ref{fig:realdata}.  Possible reasons for this discrepancy were given in (\ref{prob:2ngon}) and (\ref{prob:GKXR}). A detailed assessment, however, remains for future research.
\end{enumerate}

\subsection{Determination of faceting} \label{synthsection:disc3}
\begin{figure}[ht]
\centering
\includegraphics[width= 1.0\columnwidth]{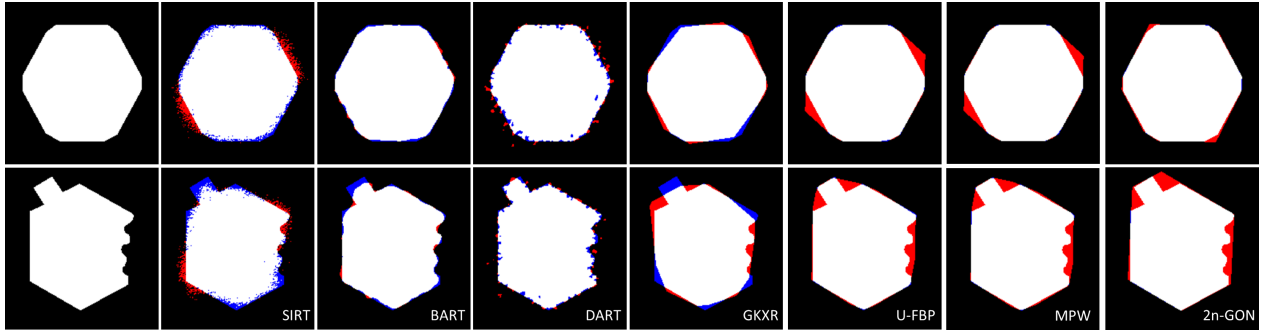}
\caption{Phantoms and reconstructions~II. Column~1: Phantoms 5 (top) and 6 (bottom). Difference images between a phantom and a typical reconstruction with $S_{140,10}$ at $50$-noise obtained by SIRT (Column~2), BART (Column~3), DART (Column~4), GKXR (Column~5), U-FBP (Column~6), MPW (Column~7), and $2n$-GON (Column~8).  Color scheme for difference images: White  pixels belong to the phantom (P) and the reconstruction (R), red pixels to $R\setminus P$, and blue  pixels to $P\setminus R$. These are $192\times192$ pixel images in which the $160$ pixel boundary has been cropped. \label{fig:fig11}}
\end{figure}

A typical experimental aim when imaging geometric objects is to determine their faceting including minor facets and surface protrusion and roughness. Fig.~\ref{fig:fig11} shows two more complex phantoms along with reconstructions obtained by SIRT, BART, DART, GKXR, U-FBP, MPW, and $2n$-GON, respectively. We refer to these as Phantoms~5 and~6; reconstructions for demonstration purposes have been performed for $S_{140,10}$ at $50$-noise (the data generation was described in Section~\ref{sect:datageneration}).

The reconstruction errors $(\delta_S;\delta_H)$ for Phantoms~5 and 6 are, respectively: For SIRT, $(1,501;13)$ and $(1,742;12)$; for BART, $(460;3)$ and $(692;9)$; for DART, $(842;11)$ and $(769;14)$; for GKXR, $(934;5)$ and $(1,744;11)$; for U-FBP, $(852;10)$ and $(1,199;8;)$; for MPW, $(791;10)$ and $(1,243;8)$; and for $2n$-GON, $(379;5)$, and $(1,443;12)$.

The results in Fig.~\ref{fig:fig11} can be summarized as follows.
\begin{itemize}
\item SIRT: Minor facets of Phantom~6 are not reconstructed; the boundaries are very fuzzy; additional imaging tools for edge extraction are required to evaluate facet angles; the inner angles of the major facets are reconstructed with an accuracy of about $12$ degrees.
\item BART: All minor facets are reconstructed; artefacts in the missing wedge region might be misclassified as minor facets; additional imaging tools for edge extraction are required to evaluate facet angles; the inner angles of the major facets are reconstructed with an accuracy of about $2$ degrees.
\item DART: Several minor facets are reconstructed; a few minor facets in the missing wedge region are not resolved; boundaries can be fuzzy and additional imaging tools for edge extraction are required to evaluate facet angles; assuming that only isolated pixels are removed, we find that the inner angles of the major facets are reconstructed with an accuracy of about $6$ degrees.
\item GKXR: Few minor facets are reconstructed; the inner angles of the major facets are reconstructed with an accuracy of about $8$ degrees (cf. Fig.~\ref{fig:fig7}).
\item U-FBP and MPW: Minor facets of Phantom~6 are not reconstructed; the inner angles of the major facets outside the missing wedge are reconstructed with an accuracy of about $2$ degrees; artificial major facets appear in the missing wedge of Phantom~5.
\item $2n$-GON: Minor facets of Phantom~6 and two minor facets of Phantom~5, are not reconstructed; the inner angles of the major facets are reconstructed with an accuracy of about $1.5$ degrees.
\end{itemize}
These results demonstrate the potential of the geometric approach. A detailed study, however, must be left for future research, since the reconstruction quality depends on several parameters, such as the particular type of noise, the misalignment of the projections, and the relative position of the object and the missing wedge.

As a final comment, we remark that a large $L_2$-norm of the residual $Ax-b$ can be seen as an indicator that the reconstruction cannot be trusted (notation as in Section~\ref{sect:SIRT}). For the GKXR, MPW, and $2n$-GON reconstructions in Fig.~\ref{fig:fig11}, for instance, we find that $||Ax-b||$ is at least three times larger for Phantom~6 than for Phantom~5.

\section{Conclusion}
We have applied five algorithms from the mathematical fields of geometric and discrete tomography to reconstruct homogeneous objects from electron tomography data. Our results demonstrate that the choice of reconstruction algorithm should be based on the specific reconstruction task at hand.  None of the algorithms considered is better for all reconstruction tasks than the others. The main features of the algorithms that we introduced here in the ET context can be summarized as follows: BART and DART reconstruct homogeneous objects from projections; GKXR reconstructs convex objects from only four projections; MPW reconstructs convex objects from shadows and compensates for noise; and $2n$-GON reconstructs objects that are close to regular $2n$-gons from (possibly few) noisy shadows.

\section{Acknowledgments} The first and third author were supported by DFG grant AL 1431/1-1 and GR 993/10-1, respectively.  The second author was supported in part by NSF grants DMS-0603307 and DMS-1103612. The eighth author acknowledges financial support by the NWO (the Netherlands Organisation for Scientific Research), project number 639.072.005. We thank Gabor T. Herman for valuable discussions related to SNARK09 and BART. The GKXR algorithm, with simulated projections of convex polygons and ellipses, was implemented by Mark Lockwood and modified to accept real-world data by Kyle Rader, both working while undergraduates at Western Washington University. We thank L. Fr\"oberg for growing and providing the nanowire specimen. Ran Davidi and Michael Ritter are acknowledged for technical support.

\appendix
\section{SNARK09 commands for BART}

See Table~\ref{table:BART}.

\setcounter{table}{0}
\begin{table}[bt]
\begin{center}
\begin{tabular}{l|l}
\begin{minipage}{0.4\columnwidth}
{\footnotesize
\begin{verbatim}
PICTURE TEST
PROJECTION REAL
MODE LOWER = 0.0
MODE UPPER = 1.0
STOP ITERATION 5
EXECUTE AVERAGE ART CONTOUR
BART on hexagon image
0.5 0 1
1
ART3 relaxation constant 0.3
CONSTRAINT BART
*
END
\end{verbatim}}
\end{minipage}
 &  \hspace*{4ex}
\begin{minipage}{0.4\columnwidth}
{\footnotesize
\begin{verbatim}
STOP ITERATION 1
BASIS PIXEL
EXECUTE CONTINUE ALP1  SMOOTH
Smoothing of BART recon
2 1 1 1
1
BASIS BLOBS
EXECUTE CONTINUE ALB1  CONTOUR
Contouring of smooth BART recon
0.5 0 1 0
1
\end{verbatim}}
\end{minipage}
\end{tabular}
\end{center}
\caption{SNARK09 code (input file) for the BART algorithm. Left: main routine. Right: filtering routine. Seven copies of the filtering routine should be placed between the ``*'' and ``\texttt{END}'' of the main routine.}\label{table:BART}
\end{table}

\providecommand{\bysame}{\leavevmode\hbox to3em{\hrulefill}\thinspace}
\providecommand{\MR}{\relax\ifhmode\unskip\space\fi MR }
\providecommand{\MRhref}[2]{%
  \href{http://www.ams.org/mathscinet-getitem?mr=#1}{#2}
}
\providecommand{\href}[2]{#2}


\begin{thebibliography}{10}

\bibitem{alpersphase}
A.~Alpers, G.T. Herman, H.F. Poulsen, and S.~Schmidt, \emph{Phase retrieval for
  superposed signals from multiple binary objects}, J. Opt. Soc. Am. A
  \textbf{27} (2010), no.~9, 1927--1937.

\bibitem{apkh-06}
A.~Alpers, H.F. Poulsen, E.~Knudsen, and G.T. Herman, \emph{A discrete
  tomography algorithm for improving the quality of {3DXRD} grain maps}, J.
  Appl. Crystallogr. \textbf{39} (2006), no.~4, 582--588.

\bibitem{DART4}
S.~Bals, K.J. Batenburg, D.~Liang, O.~Lebedev, G.~van Tendeloo, A.~Aerts, J.A.
  Martens, and C.E.A. Kirschhock, \emph{Quantitative three-dimensional modeling
  of {Z}eotile through discrete electron tomography}, J. Amer. Chem. Soc.
  \textbf{131} (2009), no.~13, 4769--4773.

\bibitem{DART3}
S.~Bals, K.J. Batenburg, J.~Verbeeck, J.~Sijbers, and G.~van Tendeloo,
  \emph{Quantitative {3D} reconstruction of catalyst particles for bamboo-like
  carbon-nanotubes}, Nano Lett. \textbf{7} (2007), no.~12, 3669--3674.

\bibitem{tomobookmater}
J.~Banhart, \emph{Advanced tomographic methods in materials research and
  engineering}, Monographs on the Physics and Chemistry of Materials, Oxford
  University Press, Oxford, 2008.

\bibitem{DART1}
K.J. Batenburg, S.~Bals, J.~Sijbers, C.~Kuebel, P.A. Midgley, J.C. Hernandez,
  U.~Kaiser, E.R. Encina, E.A. Coronado, and G.~van Tendeloo, \emph{{3D}
  imaging of nanomaterials by discrete tomography}, Ultramicroscopy
  \textbf{109} (2009), no.~6, 730--740.

\bibitem{DART2}
K.J. Batenburg and J.~Sijbers, \emph{{DART}: {A} practical reconstruction
  algorithm for discrete tomography}, IEEE Trans. Image Process. \textbf{20}
  (2011), no.~9, 2542--2553.

\bibitem{graylevel1}
K.J. Batenburg, W.~van Aarle, and J.~Sijbers, \emph{A semi-automatic algorithm
  for grey level estimation in tomography}, Pattern Recogn. Lett. \textbf{32}
  (2011), no.~9, 1395--1405.

\bibitem{DART6}
E.~Biermans, L.~Molina, K.J. Batenburg, S.~Bals, and G.~van Tendeloo,
  \emph{Measuring porosity at the nanoscale by quantitative electron
  tomography}, Nano Lett. \textbf{10} (2010), no.~12, 5014--5019.

\bibitem{Gibbs1}
B.M. Carvalho, G.T. Herman, S.~Matej, C.~Salzberg, and E.~Vardi, \emph{Binary
  tomography for triplane cardiography}, Information Processing in Medical
  Imaging (A.~Kuba, M.~Samal, and A.~Todd-Pokropek, eds.), vol. 1613, Springer,
  Berlin, 1999, pp.~29--41.

\bibitem{coltonkress}
D.~Colton and R.~Kress, \emph{Inverse acoustic and electromagnetic scattering
  theory}, Applied Mathematical Sciences 93, Springer, Berlin, 1992.

\bibitem{snark09}
R.~Davidi, G.T. Herman, and J.~Klukowska, \emph{{SNARK09: A} programming system
  for the reconstruction of {2D} images from {1D} projections},
  \texttt{http://www.dig.cs.gc.cuny.edu/software/snark09}, 2009, [Online;
  accessed 10-August-2012].

\bibitem{nanogrowth}
K.A. Dick, \emph{A review of nanowire growth promoted by alloys and
  non-alloying elements with emphasis on {Au}-assisted {III-V} nanowires},
  Prog. Cryst. Growth. Charact. Mater. \textbf{54} (2008), no.~1--3, 138--173.

\bibitem{artifacts1}
P.~Ercius, M.~Weyland, D.A. Muller, and L.M. Gignac, \emph{Three-dimensional
  imaging of nanovoids in copper interconnects using incoherent bright field
  tomography}, Appl. Phys. Lett. \textbf{88} (2006), no.~24, 243116.

\bibitem{micronpaper}
J.-J. Fernandez, \emph{Computational methods for electron tomography}, Micron
  \textbf{43} (2012), no.~10, 1010--1030.

\bibitem{artifacts2}
H.~Friedrich, M.R. McCartney, and P.R. Buseck, \emph{Comparison of intensity
  distributions in tomograms from {BF TEM}, {ADF TEM}, {HAADF STEM}, and
  calculated tilt series}, Ultramicroscopy \textbf{106} (2005), no.~1, 18--27.

\bibitem{gardner}
R.J. Gardner, \emph{Geometric tomography}, 2nd ed., Encyclopedia of Mathematics
  and its Applications, vol.~58, Cambridge University Press, New York, 2006.

\bibitem{GarG97}
R.J. Gardner and P.~Gritzmann, \emph{Discrete tomography: {D}etermination of
  finite sets by {X}-rays}, Trans. Amer. Math. Soc. \textbf{349} (1997), no.~6,
  2271--2295.

\bibitem{GardnerKiderlen}
R.J. Gardner and M.~Kiderlen, \emph{A solution to {H}ammer's {X}-ray
  reconstruction problem}, Adv. Math. \textbf{214} (2007), no.~1, 323?--343.

\bibitem{gardnerkiderlen09}
\bysame, \emph{A new algorithm for 3{D} reconstruction from support functions},
  IEEE Trans. Pattern Anal. Mach. Intell. \textbf{31} (2009), no.~3, 556--562.

\bibitem{convergence}
R.J. Gardner, M.~Kiderlen, and P.~Milanfar, \emph{Convergence of algorithms for
  reconstructing convex bodies and directional measures}, Ann. Statist.
  \textbf{34} (2006), no.~3, 1331--1374.

\bibitem{GarM80}
R.J. Gardner and P.~McMullen, \emph{On {H}ammer's {X}-ray problem}, J. London
  Math. Soc. (2) \textbf{21} (1980), no.~1, 171--175.

\bibitem{SIRT}
P.~Gilbert, \emph{Iterative methods for the three-dimensional reconstruction of
  an object from projections}, J. Theoret. Biol. \textbf{36} (1972), no.~1,
  105--117.

\bibitem{tv1}
B.~Goris, W.~van~den Broek, K.J. Batenburg, H.~Heidari Mezerji, and S.~Bals,
  \emph{Electron tomography based on a total variation minimization
  reconstruction technique}, Ultramicroscopy \textbf{113} (2012), no.~0,
  120--130.

\bibitem{SIRT2}
J.~Gregor and T.~Benson, \emph{Computational analysis and improvement of
  {SIRT}}, IEEE Trans. Medical Imaging \textbf{27} (2008), no.~7, 918--924.

\bibitem{bart2}
G.T. Herman, \emph{Reconstruction of binary patterns from a few projections},
  International Computing Symposium 1973 (A.~G\"unther, B.~Levrat, and
  H.~Lipps, eds.), North-Holland Publ. Co., Amsterdam, 1974, pp.~371--378.

\bibitem{herman}
\bysame, \emph{Fundamentals of computerized tomography: Image reconstruction
  from projections}, 2nd ed., Advances in Pattern Recognition, Springer,
  London, 2009.

\bibitem{kubaherman1}
G.T. Herman and A.~Kuba (eds.), \emph{Discrete tomography: Foundations,
  algorithms, and applications}, Applied and Numerical Harmonic Analysis,
  Birkh\"auser Boston Inc., Boston, MA, 1999.

\bibitem{kubaherman2}
G.T. Herman and A.~Kuba (eds.), \emph{Advances in discrete tomography and its
  applications}, Applied and Numerical Harmonic Analysis, Birkh\"auser Boston
  Inc., Boston, MA, 2007.

\bibitem{artifacts3}
A.H. Janssen, C.-M. Yang, Y.~Wang, F.~Sch{\"u}th, A.J. Koster, and K.P.
  de~Jong, \emph{Localization of small metal (oxide) particles in {SBA}-15
  using bright-field electron tomography}, J. Phys. Chem. B \textbf{107}
  (2003), no.~38, 10552--10556.

\bibitem{kaipio}
J.~Kaipio and E.~Somersalo, \emph{Statistical inverse problems: Discretization,
  model reduction and inverse crimes}, J. Comput. Appl. Math. \textbf{198}
  (2007), no.~2, 493--504.

\bibitem{kakslaney}
A.C. Kak and M.~Slaney, \emph{Principles of computerized tomographic imaging},
  Classics in Applied Mathematics, vol.~33, Society for Industrial and Applied
  Mathematics (SIAM), Philadelphia, PA, 2001, Reprint of the 1988 original.

\bibitem{klette}
R.~Klette and A.~Rosenfeld, \emph{Digital geometry}, Morgan Kaufmann, San
  Francisco, CA, 2004.

\bibitem{DART5}
F.~Leroux, M.~Gysemans, S.~Bals, K.J. Batenburg, J.~Snauwaert, T.~Verbiest,
  C.~van Haesendonck, and G.~van Tendeloo, \emph{{3D} characterization of
  helical silver nanochains mediated by protein assemblies}, Adv. Mater.
  \textbf{22} (2010), no.~19, 2193--2197.

\bibitem{nanowireappl1}
Y.~Lia, F.~Qian, J.~Xiang, and C.M. Lieber, \emph{Nanowire electronic and
  optoelectronic devices}, Mater. Today \textbf{9} (2006), no.~10, 18--27.

\bibitem{Lyo01}
R.G. Lyons, \emph{Understanding digital signal processing}, Prentice Hall,
  Upper Saddle River, NJ, 2001.

\bibitem{blob1}
R.~Marabini, G.T. Herman, and J.M. Carazo, \emph{{3D} reconstruction in
  electron microscopy using {ART} with smooth spherically symmetric volume
  elements (blobs)}, Ultramicroscopy \textbf{72} (1998), no.~1--2, 53--65.

\bibitem{electrontomo09}
P.A. Midgley and R.E. Dunin-Borkowski, \emph{Electron tomography and holography
  in materials science}, Nature Mater. \textbf{8} (2009), no.~4, 271--280.

\bibitem{artifacts4}
P.A. Midgley and M.~Weyland, \emph{{3D} electron microscopy in the physical
  sciences: the development of {Z}-contrast and {EFTEM} tomography},
  Ultramicroscopy \textbf{96} (2003), no.~3--4, 413--431.

\bibitem{artifacts5}
G.~M{\"o}bus, R.C. Doole, and B.J. Inkson, \emph{Spectroscopic electron
  tomography}, Ultramicroscopy \textbf{96} (2003), no.~3--4, 433--451.

\bibitem{electrontomoreview}
G.~M{\"o}bus and B.J. Inkson, \emph{Nanoscale tomography in materials science},
  Mater. Today \textbf{10} (2007), no.~12, 18--25.

\bibitem{nanowireappl2}
S.~Mokkapati and C.~Jagadish, \emph{{III-V} compound {SC} for optoelectronic
  devices}, Mater. Today \textbf{12} (2009), no.~4, 22--32.

\bibitem{blob2}
T.~Obi, S.~Matej, R.M. Lewitt, and G.T. Herman, \emph{{2.5-D} simultaneous
  multislice reconstruction by series expansion methods from {F}ourier-rebinned
  {PET} data}, IEEE Trans. Medical Imaging \textbf{19} (2000), no.~5, 474--484.

\bibitem{CTalgorithms09}
X.~Pan, E.Y Sidky, and M.~Vannier, \emph{Why do commercial {CT} scanners still
  employ traditional, filtered back-projection for image reconstruction?},
  Inverse Problems \textbf{25} (2009), no.~12, 123009.

\bibitem{ourpaper}
R.S. Pennington, S.~K{\"o}nig, A.~Alpers, C.B. Boothroyd, and R.E.
  Dunin-Borkowski, \emph{Reconstruction of an {I}n{A}s nanowire using geometric
  and algebraic tomography}, J. Phys.: Conf. Ser. \textbf{326} (2011), no.~1,
  012045.

\bibitem{petersen09}
T.C. Petersen and S.P. Ringer, \emph{Electron tomography using a geometric
  surface-tangent algorithm: application to atom probe specimen morphology}, J.
  Appl. Phys. \textbf{105} (2009), no.~10, 103518.

\bibitem{princewillsky90}
J.L. Prince and A.S. Willsky, \emph{Estimating convex sets from noisy support
  line measurements}, IEEE Trans. Pattern Anal. Machine Intell. \textbf{12}
  (1990), no.~4, 377--389.

\bibitem{tv2}
Z.~Saghi, D.J. Holland, R.~Leary, A.~Falqui, G.~Bertoni, A.J. Sederman, L.F.
  Gladden, and P.A. Midgley, \emph{Three-dimensional morphology of iron oxide
  nanoparticles with reactive concave surfaces. {A} compressed sensing-electron
  tomography ({CS-ET}) approach}, Nano Lett. \textbf{11} (2011), no.~11,
  4666--4673.

\bibitem{electrontomo08}
Z.~Saghi, X.~Xu, and G.~M{\"o}bus, \emph{Electron tomography of regularly
  shaped nanostructures under non-linear image acquisition}, J. Microsc.
  \textbf{232} (2008), no.~1, 186--195.

\bibitem{graylevel2}
W.~van Aarle, K.J. Batenburg, and J.~Sijbers, \emph{Automatic parameter
  estimation for the discrete algebraic reconstruction technique (dart)}, to
  appear in IEEE Trans. Image Proc., 2012.

\bibitem{joostnature}
S.~van Aert, K.J. Batenburg, M.D. Rossell, R.~Erni, and G.~van Tendeloo,
  \emph{Three-dimensional atomic imaging of crystalline nanoparticles}, Nature
  \textbf{470} (2011), no.~7334, 374--377.

\bibitem{hexagoncrosssection}
J.B. Wagner, N.~Sk{\"o}ld, L.R. Wallenberg, and L.~Samuelson, \emph{Growth and
  segregation of {GaAs-Al$_x$In$_{1-x}$P} core-shell nanowires}, J. Cryst.
  Growth \textbf{312} (2010), no.~10, 1755--1760.

\bibitem{artifacts6}
X.~Xu, Y.~Peng, Z.~Saghi, R.~Gay, B.J. Inkson, and G.~M{\"o}bus, \emph{{3D}
  reconstruction of {SPM} probes by electron tomography}, J. Phys.: Conf. Ser.
  \textbf{61} (2007), 810--814.

\end{thebibliography}

\end{document}